\documentclass[
preprint,
a4paper,
12pt,
superscriptaddress,
pre]{revtex4}
\usepackage{latexsym} 
\usepackage{amsmath} 
\usepackage{setspace}
\usepackage{color}
\usepackage{hyperref}
\usepackage{fancyhdr}
\usepackage{graphicx}
\usepackage{wrapfig}

\usepackage[margin=2cm]{geometry}

\begin{document} 

\title{Combined collapse by bridging and self-adhesion in a
  prototypical polymer model inspired by the bacterial nucleoid.}

\author{Vittore~F. Scolari}
\affiliation{Sorbonne Universit\'es, UPMC Univ Paris 06, UMR 7238,
  Computational and Quantitative Biology, 15 rue de l'\'{E}cole de
  M\'{e}decine Paris, France}
\affiliation{CNRS, UMR 7238, Paris, France}
\affiliation{National Centre for Biological Sciences, Tata Institute
  of Fundamental Research, GKVK, Bellary Road, Bangalore 560065,
  India}
\affiliation{Manipal University, Manipal 576104, India}
%
%
\author{Marco Cosentino Lagomarsino}
\affiliation{Sorbonne Universit\'es, UPMC Univ Paris 06, UMR 7238,
  Computational and Quantitative Biology, 15 rue de l'\'{E}cole de
  M\'{e}decine Paris, France}
\affiliation{CNRS, UMR 7238, Paris, France}

\begin{abstract}
  Recent experimental results suggest that the \emph{E.~coli}
  chromosome feels a self-attracting interaction of osmotic origin,
  and is condensed in foci by bridging interactions.
  Motivated by these findings, we explore a generic modeling framework
  combining solely these two ingredients, in order to characterize
  their joint effects. Specifically, we study a simple polymer physics
  computational model with weak ubiquitous short-ranged self
  attraction and stronger sparse bridging interactions.  Combining
  theoretical arguments and simulations, we study the general
  phenomenology of polymer collapse induced by these dual
  contributions, in the case of regularly-spaced bridging.
  Our results distinguish a regime of classical Flory-like
  coil-globule collapse dictated by the interplay of excluded volume
  and attractive energy and a switch-like collapse where bridging
  interactions compete with entropy loss terms from the looped arms of
  a star-like rosette.
  Additionally, we show that bridging can induce stable
  compartmentalized domains. In these configurations, different
  ``cores'' of bridging proteins are kept separated by star-like
  polymer loops in an entropically favorable multi-domain
  configuration, with a mechanism that parallels micellar polysoaps.
  Such compartmentalized domains are stable, and do not need any
  intra-specific interactions driving their segregation.  Domains can
  be stable also in presence of uniform attraction, as long as the
  uniform collapse is above its theta point.
\end{abstract}

\maketitle

\setstretch{1}

\section{Introduction}

It is now clear that bacterial chromosomes (which exist in the cell in
a mesoscopic dynamic complex composed of DNA, RNA and proteins called
nucleoid) are highly organized within cells. The conformational
properties of the folded genome are essential for the processes of
replication, transcription (and thus regulation of gene expression),
and segregation~\cite{Benza2012,Dillon2010,Muskhelishvili2010}.

Focusing on \emph{E.~coli}, the chromosome is a single circular
molecule of about $4.7$ million base pairs (Mbp) ($\approx1.5$
mm)~\cite{Trun1998,Stavans2006}. Nucleoid associated proteins, or
``NAPs'' (such as Dps and transcription factors Fis, H-NS, IHF, HU,
and condensin MukBEF), can modify the shape of the DNA both at local
and global levels~\cite{Dillon2010,LNW+06,Ohniwa2011}.
Of particular interest are bridging interactions~\cite{Wiggins2009}
(possible at least from Fis, H-NS, and MukBEF), which can in principle
induce looped domain formation, through mechanisms that are believed
to be important also for eukaryotic
chromatin~\cite{Brackley2013,Barbieri2013b,Junier2010}.  For example,
a study combining super-resolution microscopy with genetic
``chromosome-conformation capture'' (3C) techniques on the NAP H-NS
explicitly reported it to form a small set of foci in the cell,
bringing together distant binding sites~\cite{Wang2011a}. Additionally
H-NS
reduces the size of purified nucleoids~\cite{Thacker2013}.
RNA polymerase, the DNA-binding enzyme responsible for gene
transcription, might also concentrate into transcription foci or
``factories,'' affecting the nucleoid structure by bringing together
distant loci~\cite{JC06,Grainger2005}.

The \emph{E.~coli} nucleoid, with a linear size of $1.5$mm, occupies a
well-defined region of the cell, with a volume of 0.1-0.2
$\mu$m$^3$ (the bare DNA volume is about a factor 20-30
smaller)~\cite{Stavans2006}.  Strong nucleoid compaction into a
structure that does not fill the volume of the cell is experimentally
observed \emph{in vivo}~\cite{Zim06b-a,HadizadehYazdi2012}. %
The degree of compaction is modulated by the cell's growth conditions and
in response to specific external cues.  Rather than confinement from
the cell boundaries, the dominant force for this compaction is likely
to come from self-attraction due to molecular crowding and forces of
entropic origin effectively causing a short-ranged
self-attraction~\cite{Odi98,Vries2010}. This self-adherent polymer
organization is consistent with both \emph{in vivo}
observations~\cite{HadizadehYazdi2012,Fisher2013} and \emph{in vitro}
experiments~\cite{Pelletier2012} with purified nucleoids.
Note that compaction from bridging alone is not likely to be
responsible of this behavior, as cytoplasm-free nucleoids are larger
than cells, even if some NAPs are reported to stay
bound~\cite{Pelletier2012,Wegner2012,Thacker2013}.
Finally, a sub-Rouse viscoelastic dynamics of individual loci, whose
mean apparent diffusion varies with chromosomal
coordinates~\cite{Javer2013,Weber2010} suggests that (i) a simple
polymer model is not likely to fully capture nucleoid organization
(ii) the organization and dynamics of inter-loci tethering might also
be complex.

Importantly, the state of the nucleoid is far from being an amorphous
mass, randomly organized, such as e.g., one expects from a classical
equilibrium collapsed globule~\cite{Mirny2011}. On the contrary, the
picture of a folded object with persistent (but dynamic) mesoscopic
features, including a linear ordering of loci within the cell an
overall coiled shape should be more
realistic~\cite{WLP+06,Wiggins2010,HadizadehYazdi2012,Fisher2013,Pelletier2012}.
In these respects, one important reported feature of the
\emph{E.~coli} chromosomes are the so-called
``macrodomains''~\cite{Valens2004,MRB05,EB06}. Often described as isolated
compartments, such domains are roughly replichore-symmetric, i.e.,
mirror the order of replication of the \emph{E.~coli} genome, from the
replication origin locus, oriC, to the terminus region, Ter). The
first evidence for macrodomains~\cite{Valens2004} came from a non uniform
pattern in the recombination frequency between chromosomal loci (which
should be proportional to the population-averaged probability that the
two chromosomal segments come into contact within the cell).
Four macrodomains of a few hundred Kb in size have been identified,
which divide the chromosome into six contiguous
regions~\cite{Dame2011,Benza2012}.  Subsequent studies have confirmed
the presence of macrodomains using fluorescently labeled
loci~\cite{EMB08,EB06,LMB+05}. The Ter macrodomain appears to be
condensed by a single DNA-binding NAP, MatP, which has a small set
of specific binding sites in the Ter region~\cite{Mercier2008}.
A recent modeling study has implicated the differential condensation
levels by macrodomains, together with the targeting of the Ori and Ter
regions to specific subcellular positions, to the generation of the
chromosome segregation pattern observed \emph{in
  vivo}~\cite{Junier2013}.

Additionally, nucleoids are composed of topologically unlinked dynamic
domain structures, due to supercoiling (torsional constraints
generated by active processes and frozen by bridging) forming
plectonemes and toroids~\cite{Trun1998}, and stabilized by
nucleoid-associated proteins, such as Fis and H-NS. This combination
of effects gives the chromosome a looped
shape~\cite{Postow2004,Skoko2006,Kavenoff1976}, where the loops form a
tree of plectonemes. Supercoiling and nucleoid organization affect
gene expression~\cite{Breier2004,Postow2004,Dillon2010}.
The level of supercoiling is tightly regulated by the cell, and it can
be changed by the action of specific enzymes such as topoisomerases
and gyrases.

Sequencing techniques (though relying on population averages) give
further insight into the folding of bacterial chromosomes.
High-throughput 3C techniques have been used to determine the global
folding architecture of the \emph{C.~crescentus} swarmer cell
genome~\cite{Le2013,Umbarger2011}, which is easier to access
experimentally than \emph{E.~coli}, due to the well-characterized
polar tethering of the chromosome and
the more practicability of cell synchronization.
These data show a chromosomal fiber-like organization, linearly
ordered in a compressed ring-like fiber, and taking an eight shape
inside the cell.
Higher-resolution data~\cite{Le2013} also show spatial domains of
interacting, and exhibit a hierarchical nested organization over a
range of length scales ($\sim50-200$ Kb). These domains are stable
throughout the cell cycle and are reestablished concomitantly with DNA
replication.  Additionally, domain boundaries co-occur with
highly-expressed genes, and the domain organization is enabled by
transcription. Such domains are hypothesized to be composed of
transcription-induced supercoiled plectonemes arrayed into a ``bottle
brush'' fiber.
Regarding \emph{E.~coli}, the current resolution appears too
low~\cite{Cagliero2013} to draw any specific conclusions, but
finer-scale experiments are expected to appear soon.

While the bridging and loop-forming interactions are sometimes
incorporated in polymer models of the bacterial
chromosome~\cite{Barbieri2013a,Junier2013,Fritsche2012,Heermann2012,Barbieri2012,Junier2010},
the standard approach is to neglect a possible self-adhesion and
consider a confined polymer within the cell volume.
Here, we set out to investigate how the combination of bridging and
homogeneous collapse in a generic polymer physics (equilibrium)
framework, using computer simulations and with the help of theoretical
mean-field and scaling arguments.
Rather than intending this as an explicit model for the chromosome,
our intent is to generically explore the consequences of these two
basic ingredients, in order to help more realistic model development
and to link the phenomenology with the vast existing knowledge in
polymer physics, and with chromosome models available in the
literature~\cite{Barbieri2013a,Junier2010,Benza2012}.
In this spirit, we deliberately ignore the role of supercoiling, and
we do not consider confinement. We also neglect segregation dynamics
and other non-equilibrium drives (see the Discussion).
Our main results are a qualitative characterization of the state
diagram of this model, where we observe a sharp collapse to a
rosette-like state where bridging interactions overcome entropy loss
by loops. This collapse can be described by classical Flory-like
mean-field arguments only when the density of bridges is sufficiently
high, and homogeneous self-attraction modulates the collapse by
affecting the interactions between rosette arms. Finally we find that
bridging can form multiple rosette domains, which are stable under a
wide set of conditions and whose number is set by the model
parameters.

\section{Model}

\begin{figure}
  \centering
 \includegraphics[width=0.49\textwidth]{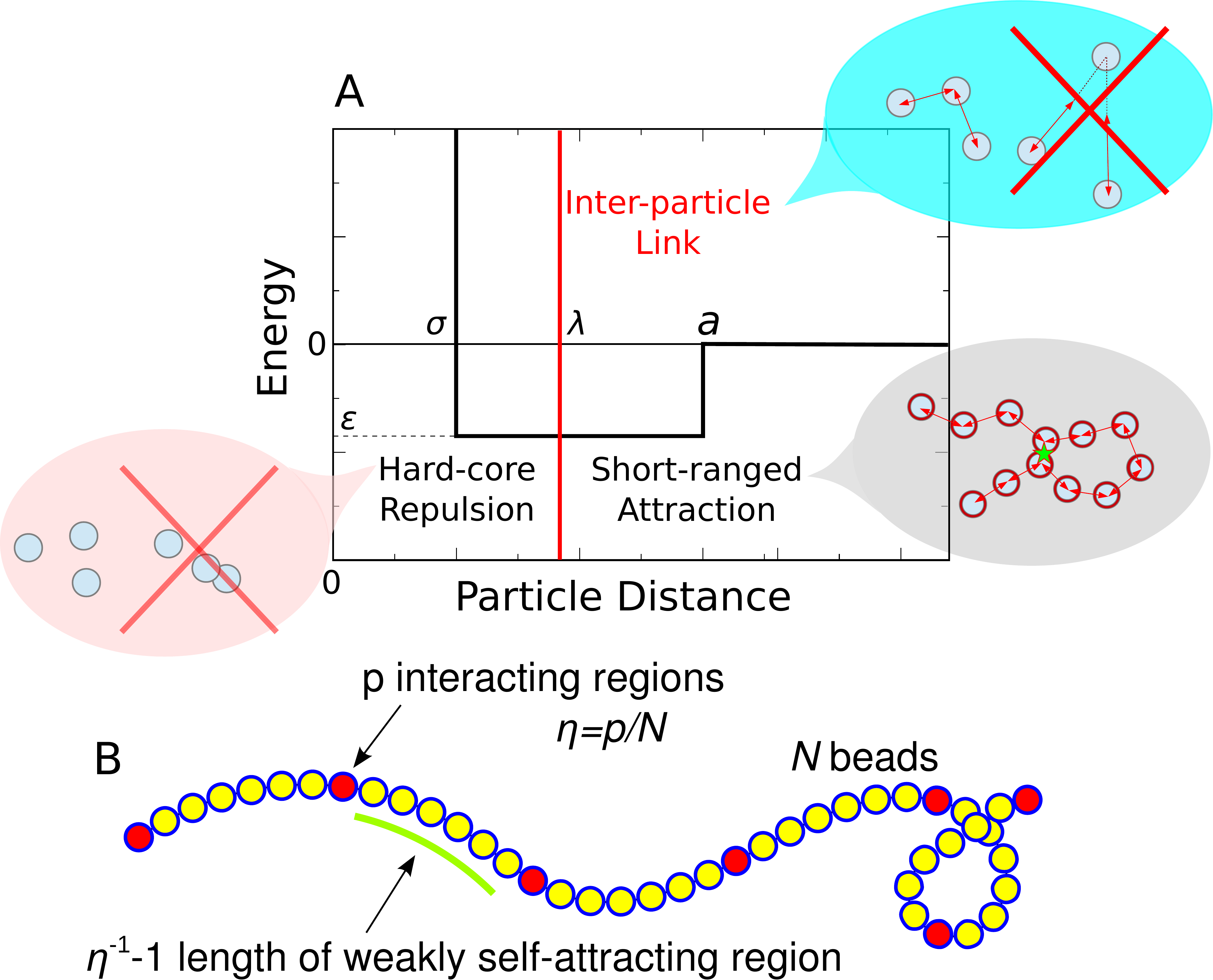}
 \caption{Illustration of the model. A: The interparticle potential
   comprises a hard-core repulsion between all discretization
   ``beads'' of the polymer (whose range is set by the parameter
   $\sigma$), and a short-ranged attraction potential of range
   $a$ and depth $\epsilon_u$ for the homogeneous self-attraction
   (acting on all beads) and $\epsilon_l$ for the sparse bridging
   interactions. Additionally, consecutive beads are subject to a
   maximum separation hard constraint of length $\lambda$.
   B: Parametrization of the position of bridging interactions. A
   total of $p$ bridging beads are placed across $p$ equally spaced
   regions of length $\simeq N /p$. The beads in the interspersing
   regions only feel the weaker self-attraction of energy
   $\epsilon_u$. }
  \label{fig:1}
\end{figure}

The basic ingredients of the model are illustrated in
Fig.~\ref{fig:1}. Our simulation uses a simple off-lattice Monte Carlo
algorithm with Metropolis rejection rule. The algorithm is a variant
of the ``bead-spring'' polymer model used in
ref.~\cite{Cacciuto2006}. The polymer is represented as a linear
string of $N$ spherical ``beads'' of diameter $\sigma$, connected by
bonds of maximal extension $\lambda \ge \sigma$.
We simulated polymers composed of up to 512 beads.  All monomers
interact via a hard-core repulsion potential
\begin{displaymath}
  U_r(r_{ij}) = \begin{cases} 
    0,  & \mbox{if }\ r_{ij} > \sigma \\ 
    \infty, & \mbox{if }\ r_{ij} \leq \sigma \ . 
  \end{cases} 
\end{displaymath}
Additionally, consecutive monomers feel the nearest-neighbor bonds as 
\begin{displaymath}
  U_b(r_{i,i+1}) = \begin{cases} 
    \infty,  & \mbox{if }\ r_{i,i+1} > \lambda \\ 
    0, & \mbox{if }\ r_{i,i+1} \leq \lambda \ . 
  \end{cases} 
\end{displaymath}
The short-ranged attraction, applied between all monomers, is modeled
as a negative square well between the two bounds imposed by the two
above potentials, and within a maximum range of $a = 1.44 \sigma$
(Fig~\ref{fig:1}A). The depth of the attractive potential is
$\epsilon_u$ for all beads, modeling a generic short-ranged attraction
due to depletion effects / molecular
crowding~\cite{Noro2000}. Bridging interactions are modeled as
sparsely chosen beads with additional square-well attractive
potentials of the same range, but acting only on other beads of the
same class with interaction energy $\epsilon_l$.
Fig.~\ref{fig:1}B illustrates the criteria for placing the bridging
interactions and their parametrization. We considered a situation
where bridging beads are equally spaced, in number $p$;  $\eta=p/N$ is
the total fraction of beads occupied by the bridging regions. We are
especially interested in the regime where $p$ is fairly small (note
that in this regime the value of $p$ may be explicitly relevant, and
consequently we will not refer to $\eta$ alone as the control
parameter). 

\section{Results}


\subsection*{Collapse from homogeneous self-attraction. }


\begin{figure}
  \centering
  \includegraphics[width=0.4\textwidth]{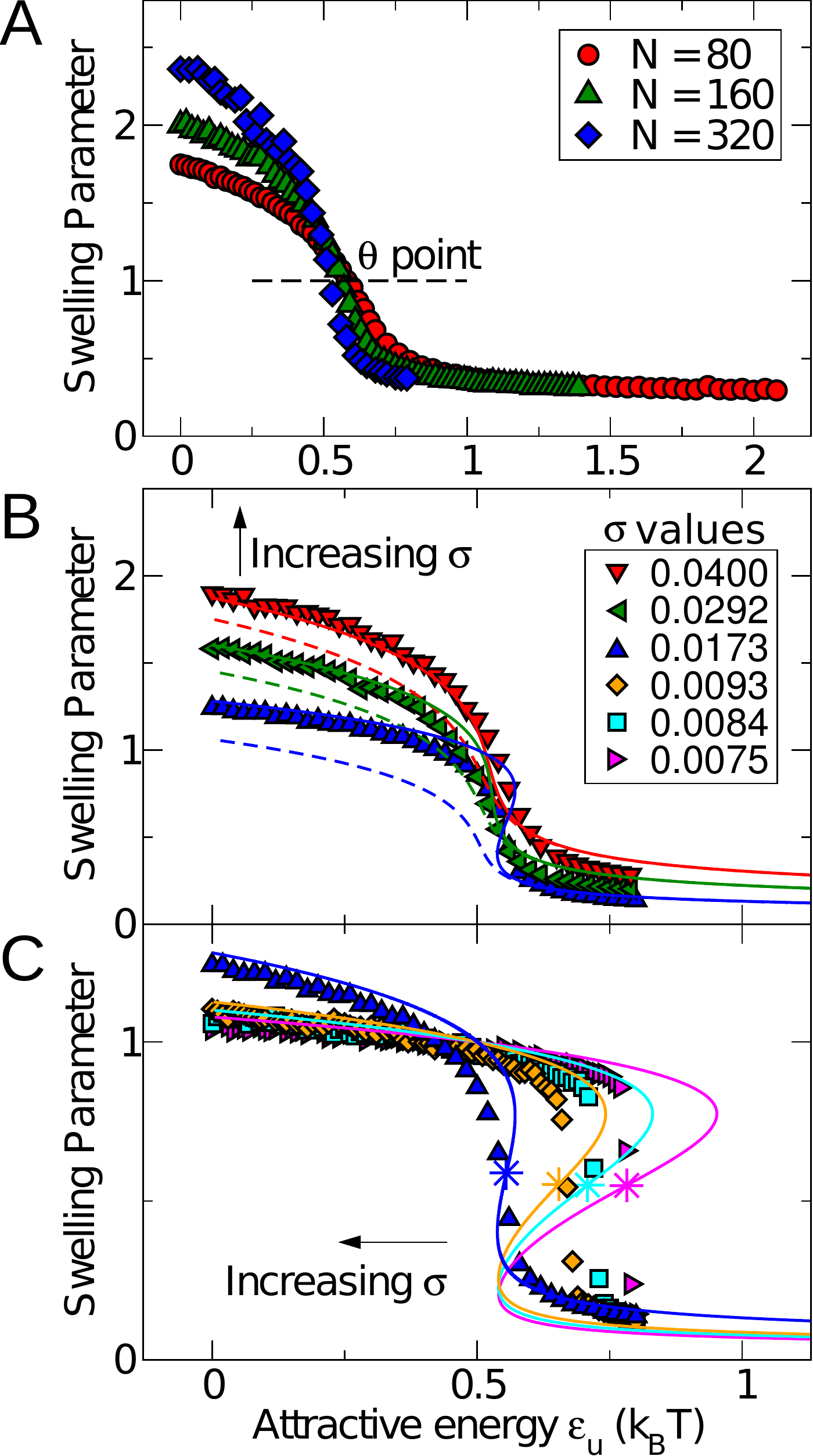}
  \caption{Collapse under homogeneous short-ranged attraction and
    Flory mean-field theory. A: Collapse curve of the swelling
    parameter $\alpha$, plotted as a function of the attraction energy
    $\epsilon_u$, for different values of $N$. In order to keep the
    swelling parameter constant in the collapsed phase, we set $\sigma
    = 0.29 N^{1/6} \lambda$ inthese simulations.  B and C: Comparison
    of collapse curves with mean-field theory for polymers with
    different excluded volume (varying $\sigma/\lambda$) at fixed $N=320$.
    Solid lines are solutions of the modified mean-field theory,
    Eq.~(\ref{eq:deg_mfield}), while dashed lines are solutions of the
    Flory mean-field theory. Panel C shows that for smaller values of
    $\sigma/\lambda$ the inflexion point of the collapse curve becomes very
    steep, and moves towards larger values of $\epsilon_u$
    energy. This feature is qualitatively captured by the modified
    Flory mean-field theory. The numerical collapse points matched the
    reentrant inflexion points of the theoretical curves, indicated by
    stars on the solid lines in the plot. }
  \label{fig:2}
\end{figure}

As a test scenario of the simulation, we first considered the limit
case of collapse of a simple polymer with homogeneous self-attraction
(a homopolymer). In this case one expects to find the conventional
theta transition, and this is the case in our simulations. To show
this, we considered the swelling parameter $\alpha$, defined here as
the ratio of the mean end-to-end distance of the polymer $R_e=\langle
|\mathbf{r}_N - \mathbf{r}_1| \rangle$ in a given condition and its
value $R_0= \lambda \left( \frac{3}{5} N \right)^{1/2}$
at the theta point.  Fig.~\ref{fig:2}A shows a plot of this quantity
as a function of the homogeneous attraction energy per bead
$\epsilon_u$, for polymers with increasing $N$. The theta point is
located where the swelling parameter equals one.  We verified that its
position corresponds well to the prediction of the Flory mean-field
theory, which defines the theta point from the balancing, in the
second virial coefficient, of the excluded volume interaction term
with the attraction term. In our case both terms can be estimated from
the Mayer function, respectively as $v_1 = \frac{4}{3} \pi \sigma^3$
(repulsion) and $v_2=  \frac{4}{3} \pi \beta \epsilon_u (a^3 - \sigma^3)$
(attraction), with $\beta=1/(k_B T)$.

In our simulations, the shape of the collapse curves of the swelling
parameter changes with varying excluded volume, i.e. varying $\sigma$
(at constant $\lambda>\sigma$).  As $\sigma$ increases, the theta point
correctly shifts towards larger energy values (Fig.~\ref{fig:2}B and
\ref{fig:2}C).
However, for increasingly ``thin'' polymers, while the theta point
still follows the predicted behavior, the inflexion point of the
swelling ratio plotted as a function of $\epsilon_u$ becomes very
steep, and radically moves towards increasingly larger values of
$\epsilon_u$ instead of smaller ones. These values, and not the theta
point, correspond to where the ``collapse'', intended as a major jump
in the swelling ratio, effectively takes place.
This phenomenology has been reported previously~\cite{DeGennes1975},
and may correspond to a first-order transition related to polymer
crystallization~\cite{Taylor2009}.  We report this in order to avoid
confusion between this sharp collapse and one of our main results,
where a sharp transition is due to bridging only. In the following we
will restrict ourselves to parameter values $\sigma = 0.424\, \lambda$
for $N=256$, where this kind of sharp collapse phenomenology due to
small $\sigma/\lambda$ \emph{does not} occur.

%
The commonly used way to find the globule size and reproduce the
collapse curve by a Flory-like mean-field theory is to counterbalance
the two-body interaction term described above with a three-body
excluded volume term. We find improved agreement with the following
variant~\cite{DeGennes1975}
  \begin{equation}
    \beta F = 3 [\alpha^2/2 - log \alpha] + 3N/5 [v \rho  + B_3 \rho^2 ] \ ,
    \label{eq:deg_mfield}
\end{equation}
where $\rho = N/R^3$ is the concentration, $v = v_1 - v_2$ plays the
role of an effective ``excluded volume'' parameter and $B_3 = (v_1)^2
/ 6$. Note the logarithmic entropic term, which can be understood as
coming from the prefactor of the radial distribution of a
freely-jointed chain.
%
%
This term becomes relevant for $\alpha < 1$ and its addition captures
rather efficiently the simulation behavior described above
(Fig.~\ref{fig:2}B).  For extremely small excluded volumes,  the
reentrant inflexion points of the theoretical collapse curve appear to
match rather well (Fig.~\ref{fig:2}C) the collapse points observed
in our simulations.

\subsection*{Collapse in presence of bridging interactions}

\begin{figure}
  \centering
  \includegraphics[width=0.4\textwidth]{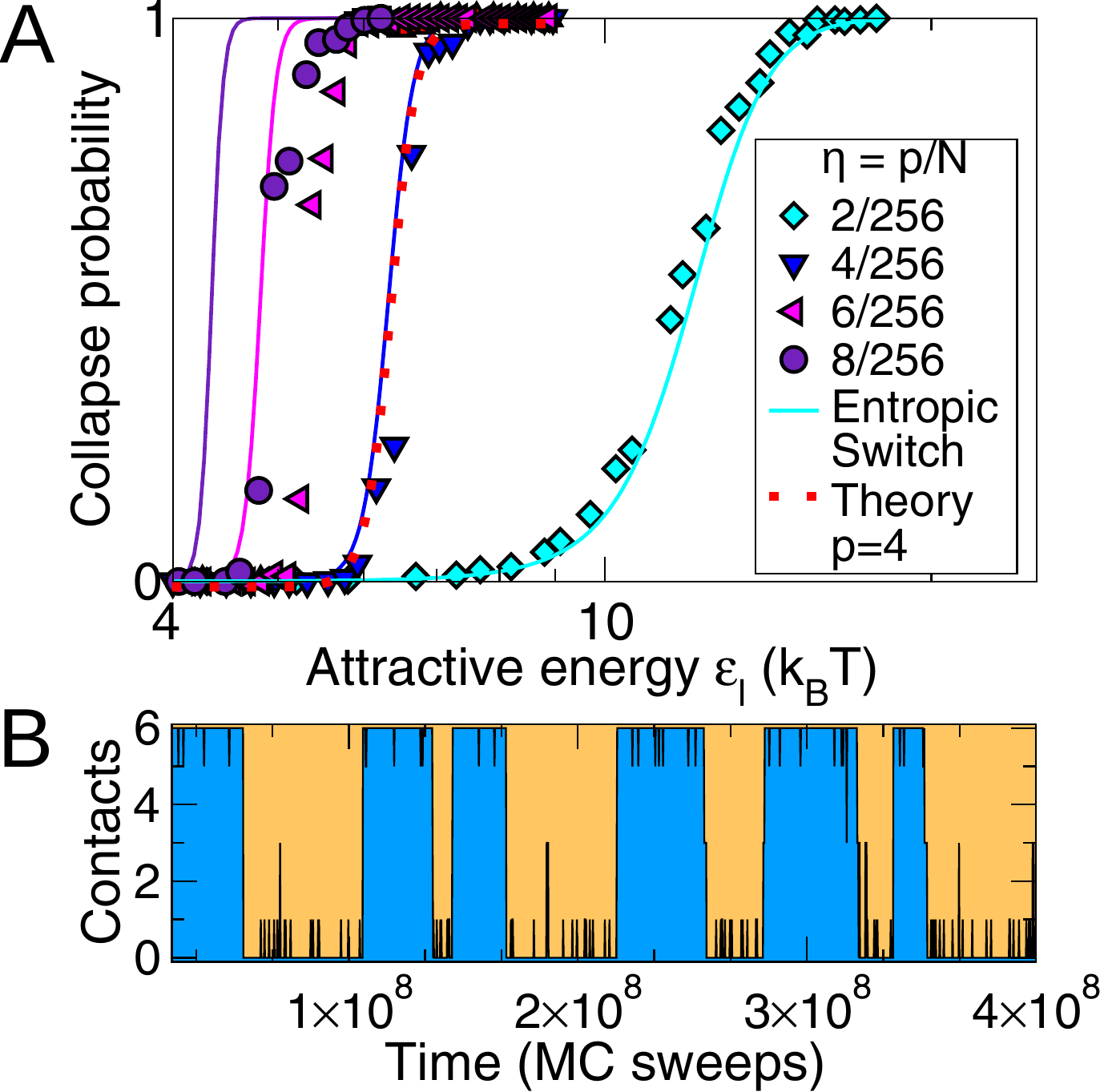}
  \caption{Collapse due to bridging proteins.  A: Switch-like
    transition to a collapsed rosette-like state for polymers for
    small numbers of bridging interactions $p$, solid lines (---) show
    the predictions of the theoretical estimates assuming two states
    (Eq.\ref{eq:Zloop_red}), while red dots
    (\textcolor{red}{$\cdots$}) correspond to the full 
    estimate for $p=4$ (Eq.~\ref{eq:prob_collapseZ}), both taking into
    account the measured scaling $P_\mathrm{(loop)} \sim
    g^{-2.27}$. The star polymer contribution has been neglected (see
    Fig.~\ref{fig:6}).
B: Two-state dynamics of the number of active bridging interactions
(contacts)  
for  $p = 4$. The interaction energy $\epsilon_l = 6.52 k_B T$ is set
close to the critical value  measured in panel A. The plot shows
switching between a compact and a swollen state as a function of Monte
Carlo time, compatibly with  a first-order phase transition showing
phase coexistence between a completely collapsed state (with six
contacts) and the swollen state.  
}
  \label{fig:3a}
\end{figure}

%

We now discuss the case of collapse driven by sparse bridging
interactions only ($\epsilon_u = 0$). This case is sometimes presented
as analogous to the classical collapse transition of a
homopolymer~\cite{Barbieri2013a,Junier2010}. While we confirm this
analogy, we also find that there are important physical differences
between the two situations. The conventional theta point and collapse
are determined essentially by a balance between excluded volume and
attractive interaction. In presence of sparse bridging points, the
relevant contribution to the partition function balanced by attractive
energy is not monomer excluded volume, but rather the entropy
reduction for closing loops between bridging
points~\cite{Marenduzzo2006c} (this kind of switch-like looping
transition has also been studied in the context of transcriptional
regulation~\cite{Saiz2006a}). In order to show this, we have
considered explicitly the following simplified partition
function~\cite{Marenduzzo2006c,Saiz2006a},
\begin{equation}
  \label{eq:Zloop}
  Z = \sum_{C} d_C e^{\beta n_b^{(C)} \epsilon_l - \Delta S(C)} \ ,
\end{equation}
where $C$ runs over the possible bridging states, with degeneracy
$d_C$
and $n_b$ is the number of bridging interactions in a given
configuration.  For small $p$ this can be estimated as the number of
pairs of interacting bridging monomers. For example, for $p=4$,
$n_b^{\mathrm{(min)}}=0$, and $n_b^{\mathrm{(max)}}=6$
(i.e. $n_b^{\mathrm{(max)}}=p(p-1)/2$).  For large cores of bridging
beads, the effects of hard core repulsion and finite-length attractive
interactions make this term linear in $p$ (proportional to the core
volume).
We suppose that there is only one ``fully collapsed'' state $C^*$
where the bridging interactions are concentrated in a spherical shaped
core (we neglect for the moment surface effects in the core, see
below). The probability for such state is
\begin{equation}
  \label{eq:prob_collapseZ}
  P(C^*) = Z^{-1}  e^{\beta n_b^{\mathrm{(max)}} \epsilon_l - \Delta
    S(C^*)} \ .
\end{equation}

To estimate the entropy loss contributions for each loop, we use the
first return probability for the polymer of length $g = (N-p)/p$. This
can be estimated for large $g$ as the return probability of a random
walk. i.e. $P_\mathrm{(loop)} \simeq (a^3/\lambda^3) g^{-3/2}$; this
approximation is correct for ghost polymers while the scaling for the
swollen state is $P_\mathrm{(loop)} \sim
g^{-2.27}$~\cite{Marenduzzo2006c}.
While the entropy of many configurations can be estimated as a
combination of loops of varying length, complex constrained
configurations can emerge that are not reducible to simple loops, but
the entropy loss terms can be computed case by case in a
straightforward way (for example for $p=4$ only one such configuration
emerges, connecting bridging points 1-3 and 2-4).  Fig.~\ref{fig:3a}A
shows a comparison of the calculation carried out for $p=4$, and
direct simulation.  The agreement between the estimated and
measured $P(C^*)$ is satisfactory, indicating that indeed a loop
entropy reduction switch is relevant. 

Additionally, we find that for $p$ sufficiently small, the dominant
contribution to the estimate is the collapsed state $C^*$, and the
transition can be captured by a simplified two-state partition
function keeping into account only the fully unfolded and the fully
bridged configurations (Fig.~\ref{fig:3a}A),
%
%
\begin{equation}
  Z_R = (p - 1)! + P_\mathrm{loop}(g)^{p-1}
            e^{\beta \epsilon_l n_b^{\mathrm{(max)}}} \ ,
  \label{eq:Zloop_red}
\end{equation}
%
%
obtained from equation \ref{eq:Zloop} and \ref{eq:prob_collapseZ} with
$\Delta S(C^*) = (p - 1)\log\left[ P_\mathrm{loop}(g)\right]$.
The prediction looses quantitative precision for increasing $p$, due
to the fact that the entropy reduction due to interactions between
loops~\cite{Hsu2004} is neglected. We verified that the estimate is
accurate for a ghost chain, and produced a refined estimate keeping
into account the star-polymer contributions at scaling level (see
Appendix~A and
Fig.~\ref{fig:6}AB). Altogether, this evidence supports the picture of
a switch-like transition to a rosette state, driven by the competition
between bridging attraction and looping entropy, and somewhat
reminiscent of microphase separation in diblock
copolymers~\cite{Leibler1980}.

\begin{figure}
  \centering
  \includegraphics[width=0.4\textwidth]{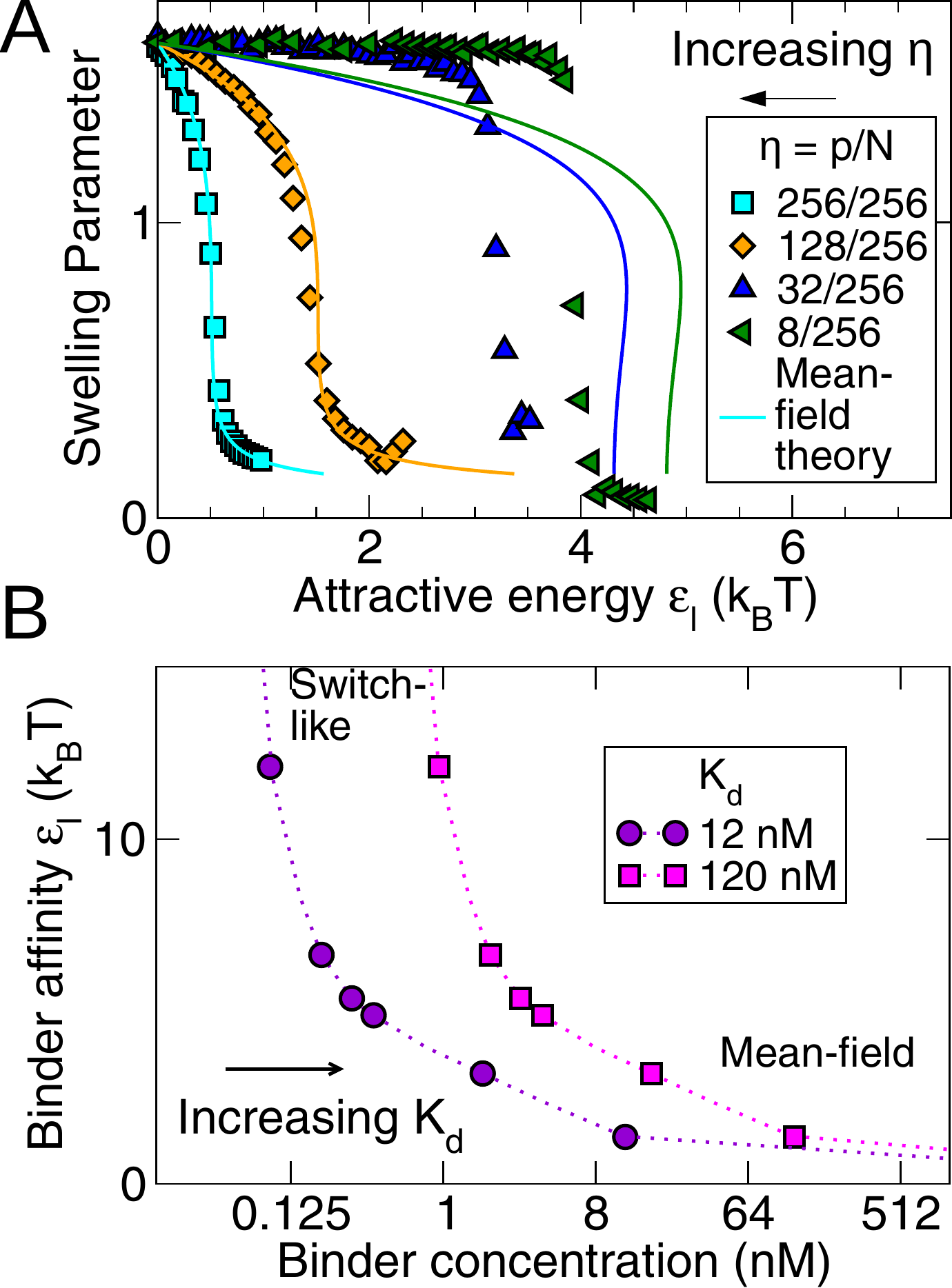}
  \caption{Collapse due to bridging proteins.
A: Collapse curves (swelling parameter vs
    $\epsilon_l$) plotted for large $p$ (and $\eta$); the transition
    energy moves to 
    lower values for decreasing $p$ (coded by color and
    symbols, see legend), the case $p/N=1$ corresponds to the case of
    uniform collapse.
The parameters have been chosen so that the phenomenology of
Fig.~\ref{fig:2}C is not present.
    The solid lines
    are fits from the Flory-like theory (Eq.~\ref{eq:deg_mfield}) on
    the parameters $\beta$ and 
    $B_3$, showing that the theory works for large $\eta= p/N$ only.
B:  Transition energy plotted as a function of an effective
concentration  of bridging proteins, estimated  from $p,N$, by a 
Langmuir model at equilibrium. We assumed different effective
dissociation constants $K_d$  
for the binders with a range of values typical for transcription
factors~\cite{Buchler29042003} ($1$ to $500$ nM). Dashed lines are
guides to the eye. The region of low binder concentration is expected
to produce a switch-like transition to a rosette state, while the
high-concentration regime is expected to follow Flory-like mean-field
behavior. An 
increase of $K_d$ corresponds to a rescaling of the transition curve. 
Simulations were performed using 256 beads and varying
$p,\epsilon_u,\epsilon_l$, with typical thermalization times of 
$1.2 \cdot 10^7$ Monte Carlo sweeps.  }
  \label{fig:3b}
\end{figure}

Fig~\ref{fig:3b}A shows the observed collapse curves for the swelling
ratio for different numbers of bridging interactions $p$ at fixed
$N$. 
We noted that, for small $p$, the transition points show better
agreement if one uses the total energy ($p \epsilon_l$) as control
parameter, rather than using $1/\eta=p/N$ as rescaling factor for
$\epsilon_l$, confirming that for small $p$, $\eta$ and $p$ are not
equivalent (in the thermodynamic limit, $p$ has to be linear in $N$ in
order to observe collapse~\cite{Kantor1996,Camacho1997}).
%
Close to the critical point, the dynamics of the swelling parameter as
a function of Monte Carlo time shows switching between two
well-defined states (Fig.~\ref{fig:3a}B), a swollen one and a compact
rosette-like one. This suggests that the switch-like transition is
likely first order (similarly to the transitions observed in diblock
copolymers).

Importantly, there is a crossover between this likely first-order and
the Flory-like (second order) homopolymer collapse behavior
(as shown in Fig.~\ref{fig:2}), when $\eta$ and $p$ are sufficiently
large.
Camacho and Schanke~\cite{Camacho1997}, working on a very similar
system, suggested the possibility of a first-order collapse transition
for sparse bridging, changing to second order for $\eta \simeq 0.6$.
We were not able in this study to explore systematically this
crossover, which resembles a standard second-order collapse for values
of $\eta = p/N$ greater than $0.3-0.4$.
However, we are not aware of existing studies fully characterizing the
crossover between the switch-like transition driven by loop entropy
and the Flory-like collapse. A less exotic possibility is that (as in
a liquid-gas phase transition) in the thermodynamic limit the
transition is always first-order, unless $\eta=1$.
%
%


Finally, Fig.~\ref{fig:3b}B makes a parallel between the case of
sparse bridging interactions with no homogeneous self-attraction, and
the ``strings and binders switch'' (SBS) model used in the context of
eukaryotic chromatin~\cite{Barbieri2012,Barbieri2013b}.
The only difference between the two situations is that in the SBS
model a pre-defined set of sparse bridging interactions can be
occupied or not by a gas of bridging binders, while the set of
bridging locations is fixed in the model considered here. To produce
an estimate mapping the two situations, we considered a Langmuir
process, and assumed it to be in chemical equilibrium.
This procedure gives a mean value of $p$ corresponding to a binder
concentration, depending on the dissociation constant $K_d$ of the
binders (we assumed $K_d$ within the range of values typical of
transcription factors~\cite{Buchler29042003}, around
$1$ to $500$ nM). Specifically, $p \simeq N / (1+K_d/c)$, where $c$ is the
concentration of binders.
We find that this rough estimate leads to a state diagram that
reproduces qualitatively the features of the SBS model
(Fig.~\ref{fig:3b}B).

\subsubsection*{Combination of sparse and homogeneous interactions} 

\begin{figure}
  \centering
  \includegraphics[width=0.4\textwidth]{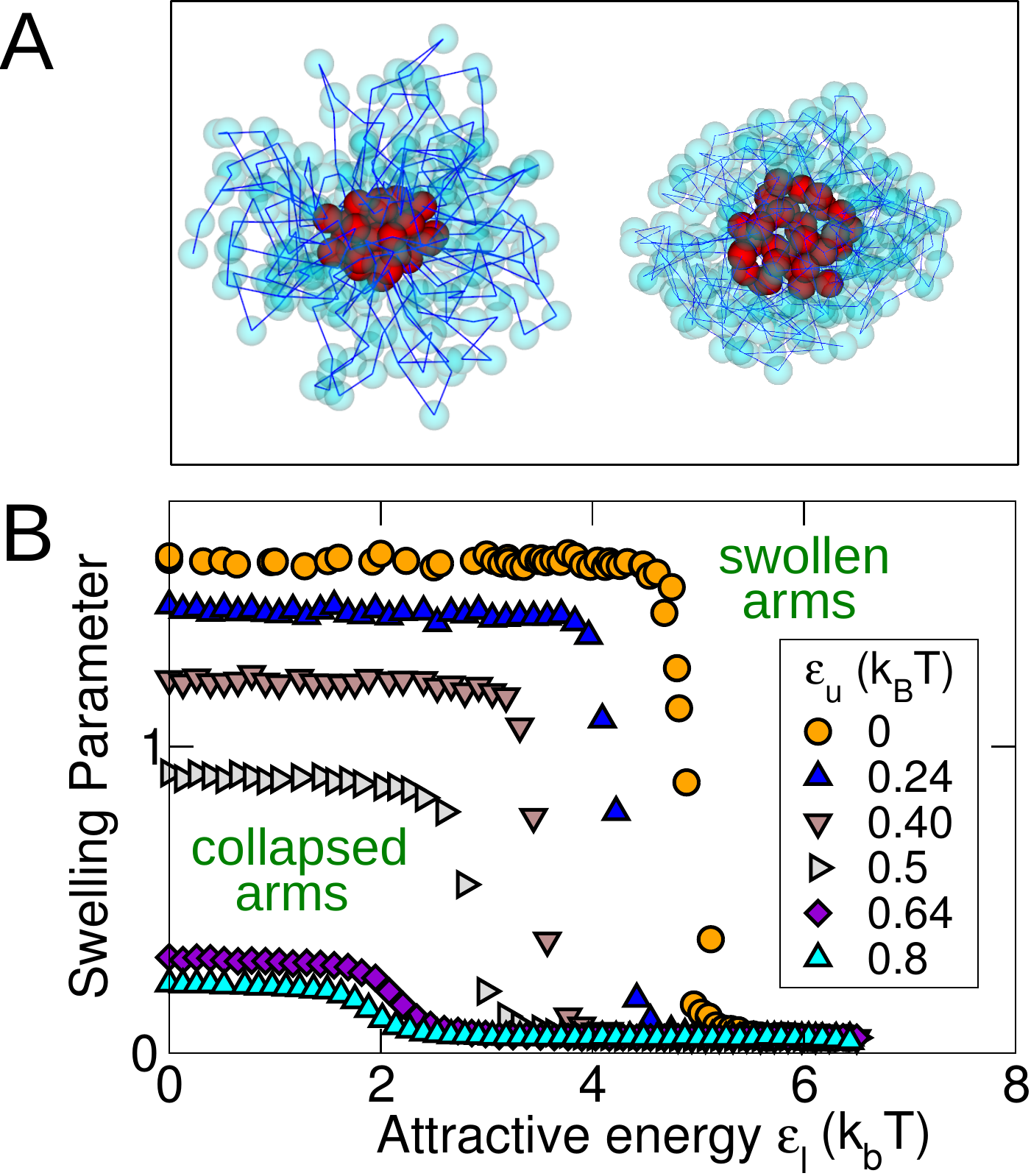}
  \caption{Combined effects of bridging interactions and
      homogeneous self-attraction.  A: Snapshots of  typical simulated
      collapsed configurations with bridging interaction stronger than
      uniform self-attraction. Left: the collapse is driven
      by bridging,  while the rest of the polymer forms closed swollen
      loops. Right: if $\epsilon_u$ is sufficiently large, both core
      of bridging beads and looped arms are collapsed. 
    B: collapse curves (swelling ratio $\alpha$ vs $\epsilon_l$) shown for
    different values of $\epsilon_u$ (coded by color and symbols, see
    legend) and $p/N = 8/256$.
    For larger values of homogeneous attraction energy $\epsilon_u$
    which put the homogeneous part of the polymer below the theta
    point, the formation of a core  of the bridging proteins still occurs inside
    the formed globule, but the point at which this  occurs
    $\left.{\epsilon_l}^*\right|_{\epsilon_u > \epsilon_u^*} \simeq 2k_BT$
    appears to depend only weakly on $\epsilon_u$.
    Simulations were performed with $N=256$ and $p=8$ for $7.5 \cdot
    10^8$ Monte Carlo sweeps, and with varying values of $\epsilon_u$
    and $\epsilon_l$ (see legend). }
  \label{fig:4}
\end{figure}

We now consider the case of collapse driven by both sparse bridging
interactions and homogeneous self-attraction ($\epsilon_u > 0$).  The
configurations of the polymer under these conditions are subject to at
least two different regimes, corresponding to the conditions in which
homogeneous interactions are large or small with respect to the
collapse energy for the homopolymer alone (we assume in both cases
that the bridging interactions are stronger $\epsilon_l >
\epsilon_u$).
Fig.~\ref{fig:4}B shows different collapse curves for the swelling
ratio of the polymer in the two regimes, for $p/N = 8/256$. In the
regime where the loops of the rosette are swollen ($v  > 0$)
the collapse is driven by the homogeneous interactions while in the
regime where the arms are collapsed ($v  < 0$) this does not
happen. 

For the swollen arms case, increasing uniform self-attraction
(i.e. increasing $\epsilon_u$) also shifts the observed transition
point $\left.{\epsilon_l}^*\right|_{\epsilon_ > \epsilon_u^*}$ towards
lower values (see Fig.~\ref{fig:4}B).  The largest contribution to
this shift comes from the reduced excluded volume interactions between
the looping chains (lower $v$). This shift reflects partially the
effect of adding the contribution of the uniform interactions directly
on the bridging beads, but we verified that this contribution is
considerably smaller.
As previously observed~\cite{Dasmahapatra2006}, the collapse curve for
increasing $\epsilon_u$ in the first regime resembles the collapse of
a homopolymer even in the presence of a small fraction of monomers
which interact more strongly. Fig.~\ref{fig:4}B shows that the effect
of bridging interactions on the swelling ratio is negligible before a
critical energy $\left.{\epsilon_l}^*\right|_{\epsilon_u >
  \epsilon_u^*} \simeq 2k_BT$, which depends on $p$, where the
bridging interactions form a core at the center of the globule. This
effect modifies the swelling ratio (defined from the end-to-end
vector) since the end beads are bridging.
%

The dependency of the free energy in the looped chains from the
excluded volume coefficient $v$ can be estimated by the analogy of the
rosette-like collapsed conformation to a star polymer with $f \simeq
p$ arms. Following Daoud and Cotton~\cite{Daoud1982} (see
Appendix~B), we estimate this entropic
contribution to the free energy to scale as
\begin{equation}
\beta F_\mathrm{(corona)} \sim f^{3/2} \log\left(Nf^{7/6} v^{1/3}
  \left( \frac{b}{r_0}\right)^{5/3} \right) \ ,
\label{eq:corona}
\end{equation}
for $v>0$ where $b$ is the monomer size and $r_0$ the size of the core. 
Inserting Eq.~\eqref{eq:corona} as additional term in the mean-field
theory leads to predict a logarithmic shift of the transition point
from $v = 0$, which is consistent with simulated data
(see Fig.~\ref{fig:4}B and Fig.~\ref{fig:6}C).

\subsection*{Compartmentalization}

\begin{figure}
  \centering
  \includegraphics[width=0.8\textwidth]{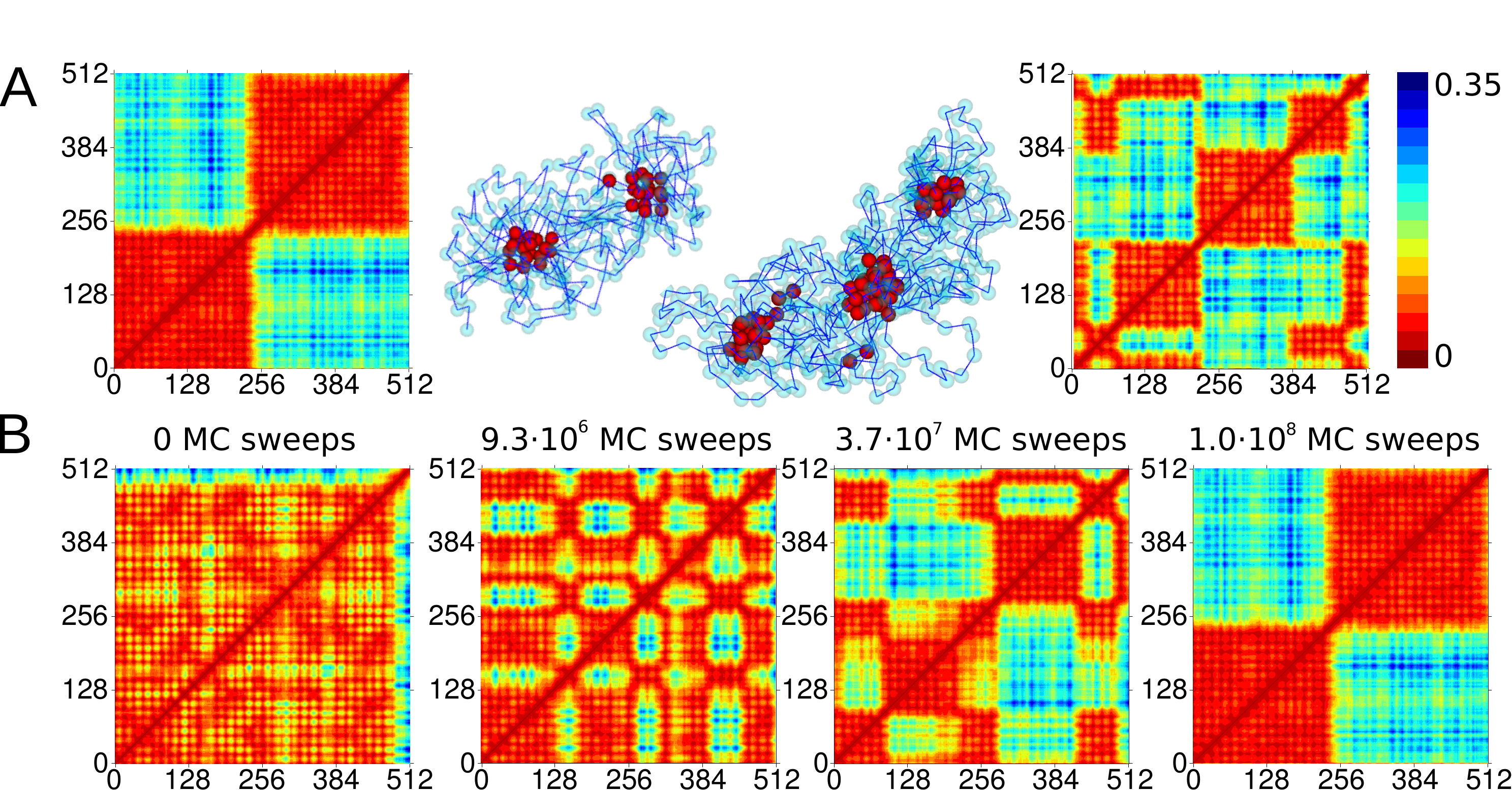}
  \caption
{Self-assembled micro-segregation into domains and shape
    anisotropy.  A: Snapshots and colormaps of mean-square distance
    between
    polymer segments (in units of $R_0^2$)
  shown  for  typical two-domain
  (left, $\epsilon_l+\epsilon_u=2.4$, $\epsilon_u=0.15$, $10^8$  Monte
  Carlo sweeps) and three-domain (right, $\epsilon_l+\epsilon_u=2.4$, 
  $\epsilon_u=0.005$, $10^7$  Monte Carlo sweeps) stable configurations;
    multiple domains are stable due the entropic repulsion of the
    loops arranged in a star-like configuration.   
B: Stability of multi-domain configurations over single-domain
collapsed state. The interaction maps of panel A are shown for
increasing Monte Carlo times. 
The  initial configuration of the simulation 
prepared as a single domain (left, as in Fig.~\ref{fig:4}A),  after
$9.3\cdot10^6$ and $3.7\cdot10^7$ Monte Carlo sweeps the system
(parameters as in panel A)  shows a
disintegration of the single-domain configuration and a progressive
reordering of the stable configuration in distinct clusters (middle), (right)
at $10^8$ sweeps a two-domain configuration is stable. 
Simulations are performed with $N = 512$, $\sigma = 0.474\, \lambda$,
$p=32$.
}
  \label{fig:5}
\end{figure}

\begin{figure}
  \centering
  \includegraphics[width=0.5\textwidth]{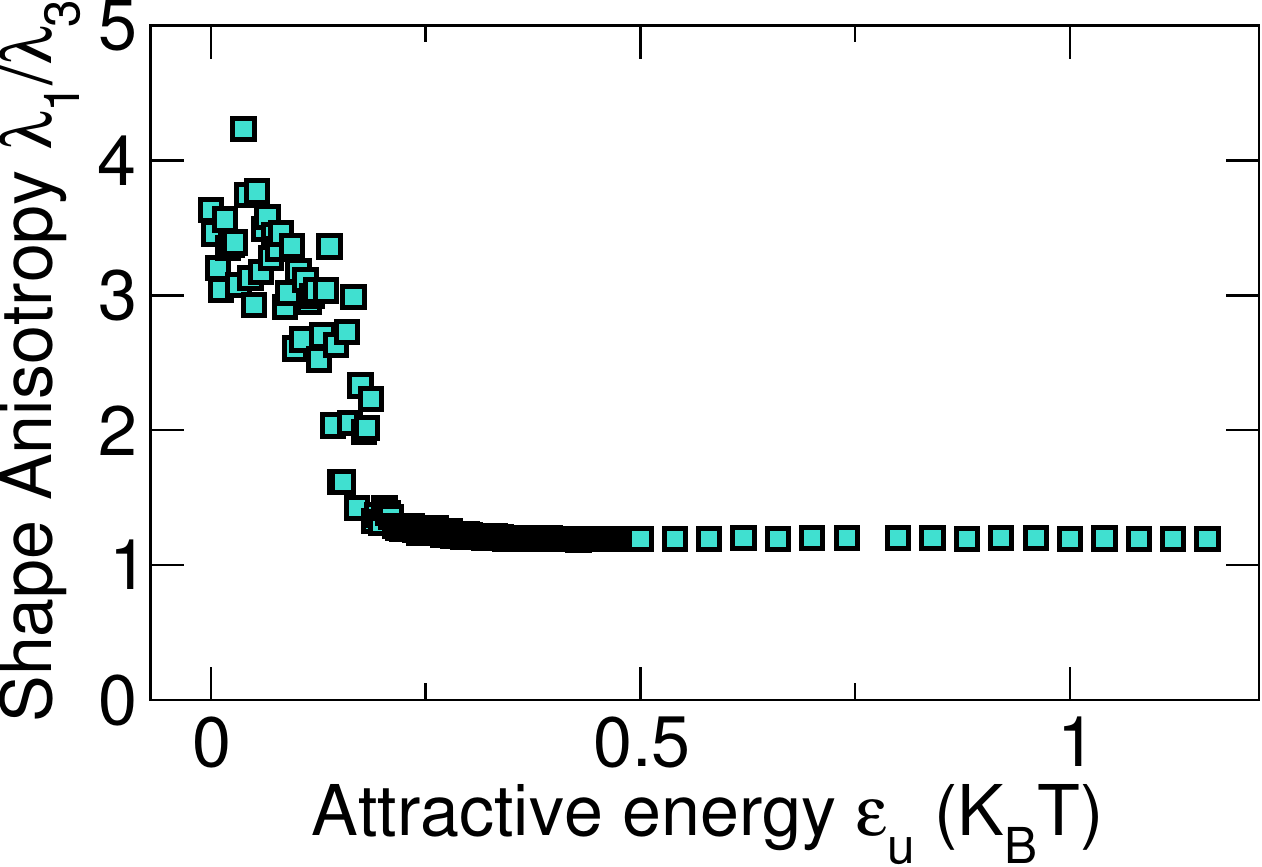}
  \caption
{
The shape anisotropy due to multiple domains (measured as the
ratio between the first and the third eigenvalue of the inertia
matrix)  shows a transition when the attractive energy $\epsilon_u$
crosses a threshold. The transition is due to coalescence of multiple
domains.  
Simulations are performed with $N = 512$, $\sigma = 0.027$ $\lambda =
0.057$, $p=32$.  }
  \label{fig:5.5}
\end{figure}

Intriguingly, we find that the switch-like collapse, in the presence
of sparse interactions can lead to long-lived states where
\emph{multiple} collapsed micelle-like domains
coexist. Fig.~\ref{fig:5}A shows snapshots of such
configurations. Multiple domains are also visible from interaction
maps, i.e. tables where the indexes are the discrete arc-length
coordinates of the polymer beads and the entries are proportional to
the mean-square distances between beads with given coordinates
(Fig.~\ref{fig:5}AB).
States with multiple domains were stable in our simulations for as
long as we could measure. Additionally, states prepared with the
initial conditions of a single domain would switch to a two- or
three-domain state, which was then observed to be stable
(Fig.~\ref{fig:5}B). This observation leads us to believe that the
multi-domain configurations might not be metastable, but true
equilibrium states.
Additionally, the states can be still observed in presence of
homogeneous self-attraction, i.e. for $\epsilon_u>0$, and the
existence of such states affects the size and the shape anisotropy of
the collapsed globules, measured as the ratio between the first and
the third eigenvalue of the polymer's inertia matrix
(Fig.~\ref{fig:5.5}). This analysis suggests that the elongated
multi-domain configurations can be stable only if $\epsilon_u$ is such
that the polymer arms are above their theta point.
%
%

In order to argue that multi-domain states could be stable, we rely on
the following scaling argument, related to the case of
polysoaps~\citep{Borisov1997,Borisov1996}. We consider a configuration
with $q$ domains, made of a ``core'' of $p/q$ bridging monomers and a
``corona'' of $p-1\simeq p/q$ loops.  Treating the loops of each
domain as a star polymer~\cite{Borisov1997} with $f=p/q$ ``arms'', and
considering the leading contribution to the entropy scaling
$f^{3/2}$~\cite{Daoud1982} leads to estimate the free energy change by
subdivision into $q$ domains as
\begin{equation}
  \Delta F_\mathrm{(corona)} \sim p^{3/2}q^{-1/2}
 \label{eq:poly1}
\end{equation}
The remaining relevant terms in the free energy are energetic, and
contain a core volume term (proportional to $\epsilon_l$) which is not
affected by the partitioning the polymer into domains, and the surface
tension term leading to the change
\begin{equation}
  \Delta F_{\mathrm{(core)}} \sim \epsilon_l (p)^{2/3}q^{1/3} \ .
 \label{eq:poly2}
\end{equation}
Minimizing the two contributions with respect to $q$ (assumed
continuous) leads to the expected equilibrium number of domains 
\begin{equation}
  q_{\mathrm{eq}} \sim p  \epsilon_l^{-6/5} \ ,
\end{equation}
indicating that the number of stable domains should increase for
larger $p$ and decrease if $\epsilon_l$ is too large. 

This behavior is observed in our simulations.  However, this argument
can be regarded as only qualitative. For example, the estimate in
Eq.~\eqref{eq:corona}, since $r_0 \sim q^{1/3}$ implies $ \Delta
F_{\mathrm{(corona)}} \sim p^{3/2}q^{-1/2} \log\left( p^{5/3}q^{-2/3}
\right)$, which (even neglecting prefactors) affects the predicted
scaling. Additionally, the argument for the entropy is consistent only
when the arms are above the theta point ($v>0$).
Note that, differently from the case of polysoaps, the multi-domain
states in this picture are due to the trade-off between the energetic
cost of the core surface when making multiple domains and the entropy
increase due to making star-like configurations with an increasing
number of arms.  Hence, according to the argument, a multi-domain
state is stable because partitioning $(p-1)$ arms into more than one
domain is less costly compared to the energy cost of partitioning the
bridging proteins in multiple domains.
Finally, for large cores, an additional contribution to the entropy
can be expected from the non-bridging beads that are ``buried'' in the
core because of their vicinity to bridging beads. The most naive
estimate for this additional cost for core size assumes that if a
bridging bead is at position $r<r_0$ within the core, it brings in
$(r_0 -r)^2$ beads (i.e. buried beads will behave like a theta-point
polymer within the core). since there are order $r^2$ beads in each
shell of the core, integrating along the radial coordinate will make
the cost scale as $r_0^5$ (i.e. $p^{5/3}$) explicitly favoring
fragmentation into multiple domains. While this effect might only need
to quantitative corrections in our simulations (since typically $p$ is
small), it might be important in realistic situations.



\section{Discussion and Conclusions. }
\label{sec:Disc}

We have analyzed a generic model of polymer collapse driven by a
combination of homogeneous and sparse attractive interactions.
Quantitative scaling arguments and simulation in parallel allowed us
to access some basic aspects of the equilibrium behavior of this
system.
There are two main results. 
First, we find a crossover between Flory-like collapse and a
switch-like, presumably first-order compaction where bridging
counterbalances loop formation. Both phenomenologies often feature as
ingredients of simple physical models for
chromatin~\cite{Marenduzzo2006c,Brackley2013,Barbieri2013b,Mirny2011,Buenemann2010},
but the interplay between the two is relatively inexplored.
Second, states with multiple micelle-like domains can exist, in a
manner that is reminiscent of micellar
polysoaps~\cite{Borisov1996}. We find that such multi-domain states
are stable: similar conclusions were reached by a recent study of a
polymer model with sparse bridging interactions, motivated by
transcription factories~\cite{Junier2010}, and motivated with a
Flory-like theory for a macroscopic extended network of foci. To
explain these structures, our work takes the complementary approach of
considering the stability of star-like rosettes.
Similar patterning (with a more complex phase diagram) has been
observed in colloidal systems driven by an external force towards a
surface with grafted polymers~\cite{C3SM50486G} or in sandwiched
polymer brushes~\cite{Curk2014}. While the phenomenology is
interesting, the latter case seems particularly interesting from the
physics viewpoint because all the main driving forces have entropic
origin.

Both of our main results are backed a series of simple
mean-field/scaling arguments.  While little of this knowledge can be
called radically new, our combined simulation and analytical approach
helps linking the results with generic knowledge in different sectors
of the polymer physics literature, including mean-field and
micelle-like collapse~\cite{DeGennes1975,Marenduzzo2006c},
polysoaps~\cite{Borisov1996}, star polymers~\cite{Daoud1982,Hsu2004}).
Overall, we believe that there is an importance in elucidating these
links, especially because they are not always kept into account in the
current landscape of models of (bacterial)
chromatin~\cite{Benza2012,Barbieri2013a}.  Interestingly, the
difference between ``homogeneous'' (Flory-like) and ``heterogeneous''
(switch-like) collapse has been explored in the 1990s with the
motivation of protein
folding~\cite{Kantor1996,Camacho1997,Bryngelson1996}. Dynamically,
they are related to the difference between ``downhill'' folding, where
large gains in stabilizing energy and loss in conformational entropy
are balanced in a way that a large range of structures can be observed
at the same time, and ``two-state'' folding, where intermediate
structures do not matter.  Biologically, one can imagine that the
collapse and swelling of genomic regions may be tuned to be
switch-like or second order, in order to be differentially controlled
externally by the cell.
Additionally, theoretical arguments are useful to elucidate the main
ingredients causing a specific effect. In particular, they lead us to
speculate that the stability of multi-micelle configurations is
conferred by the competition between surface tension of the core and
the entropic cost of the corona, which is similar to a star
polymer. Our simulation results appear to be in line with this
hypothesis. A further entropic cost due to burial of non-bridging
bonds in the core is speculated to play a role for large cores.
In analogy with diblock copolymers, one can speculate the existence of
more complex micellar phases, for example with cylindrical symmetry,
which might be exploited biologically. 
Overall, compared to previous literature, our work provides a more
precise analysis of the collapse transitions by comparing simulations
in different regimes and discusses in some more detail the orders
of transitions. Additionally, we address the role of homogeneous
self-attraction, which is usually disregarded in studies motivated by
genome organization.

The model explored here is solely based on self-adhesion and bridging,
as motivated by recent observations on bacterial chromatin.  Other
possibly important factors were voluntarily left out, in order to
obtain a cleaner description of theoretical consequences of these two
ingredients.
%
A possibly very important feature of bacterial chromosomes disregarded
by this model is the role of supercoiling, and the effectively
branched structure of plectonemes.  Modeling work on
\emph{Caulobacter}~\cite{Le2013} supports the hypothesis that
supercoil loops induced by transcription, using an effective numerical
description of supercoil-induced branching.  A more detailed model
indicates that supercoiling facilitates the probability of
protein-induced bridging~\cite{Benedetti2014}.
Regarding the specific findings reported here, on the light of these
studies we believe that a branched structure induced by supercoiling
could modulate both the loop formation entropy and the loop-loop
interaction entropy, thus affecting both the contact map and the
polymer size at fixed conditions.
%
%
Due to limitations of our simulation technique, we also left out from
this modeling framework topological constraints, which have been
implicated for eukaryotes~\cite{Mirny2011,Halverson2014}. Such
constraints lead to long-lived metastable states in a collapsing
polymer or a melt of rings, characterized by a dense fractal-like
organization (and in contrast with the dense but non-fractal
organization of an equilibrium globule). We have previously
hypothesized~\cite{Benza2012} that the peculiar sub-diffusive dynamics
of \emph{E.~coli} chromosomal loci~\cite{Weber2010,Javer2013} might be
connected to this kind of organization.

It would be premature to draw any clearcut biological conclusions
based on this simple model regarding phenomena occurring in real
chromosomes.  The main biological insights of this work are the
generic notions that the features and even nature of the collapse
transition may be tunable by modulating the sparsity of bridging and
the homogeneous self-adhesion, and that the tendency to form domains
could be intrinsic of the bridging (and tuned by the osmotic
self-adhesion), and require in principle little or no
inter-specificity of domains~\cite{Barbieri2012}. This
block-copolymer-like behavior could be generally interesting in the
context of eukaryotic chromatin. For example, mutually repellent
rosette-like chromosomal structures are also observed in some plants
and also lead to chromosome territories~\cite{Nooijer2009b}, and
bottle-brush structures are common in meiotic
chromosomes\cite{Zickler1999}.
In bacteria, this kind of spontaneous sorting mechanism might play a
role in the observed correlation between the position of genetic loci
along the chromosome and their position in the
cell~\cite{Wiggins2010,Mercier2008}, and possibly also in the
resolution of the identity of segregating sister
chromosomes~\cite{Lesterlin2012,Junier2013}.
More speculatively, the tunable transition observed here suggests
a possible more general link between the behavior of the bacterial
nucleoid to the technological area of ``smart'', or
stimulus-responsive polymers~\cite{Chen2014,Curk2014}. These are
polymer systems such as films, or polymer-colloid mixtures designed
to show a variety of responsive behaviors to external stimuli such
as light, chemicals, and solvency. Similar ``intelligent'' behavior
could be shaped into nucleoids by natural selection, and serve
biological purposes such as physiological response on fast
time-scales.

\begin{acknowledgments}
  We are very grateful to Peter Olmsted for pointing to our attention
  the literature on polysoaps, to Anton Goloborodko for suggesting the
  contribution of buried beads to the core entropy, Leonid Mirny,
  Marco Baiesi, Enrico Carlon, Mario Nicodemi, Bruno Bassetti,
  Emanuela del Gado, Bianca Sclavi, Kevin D. Dorfman and Andrew
  Spakowitz
  for discussions and useful feedback, and to Gino Benza and Ivan
  Junier
  for extremely useful comments on this manuscript.  This work was
  supported by the International Human Frontier Science Program
  Organization, grants RGY0069/2009-C and RGY0070/2014. VFS was funded
  by a PDI-MSC scholarship of the Institut de recherche pour le
  d\'eveloppement, Government of France.
\end{acknowledgments} 

\bibliography{references}

\section*{Appendix A: Role of loop interactions in the collapse due to sparse
  bridging.}
\label{sec:incl-star-polym}

\begin{figure}
  \centering
  \includegraphics[width=0.4\textwidth]{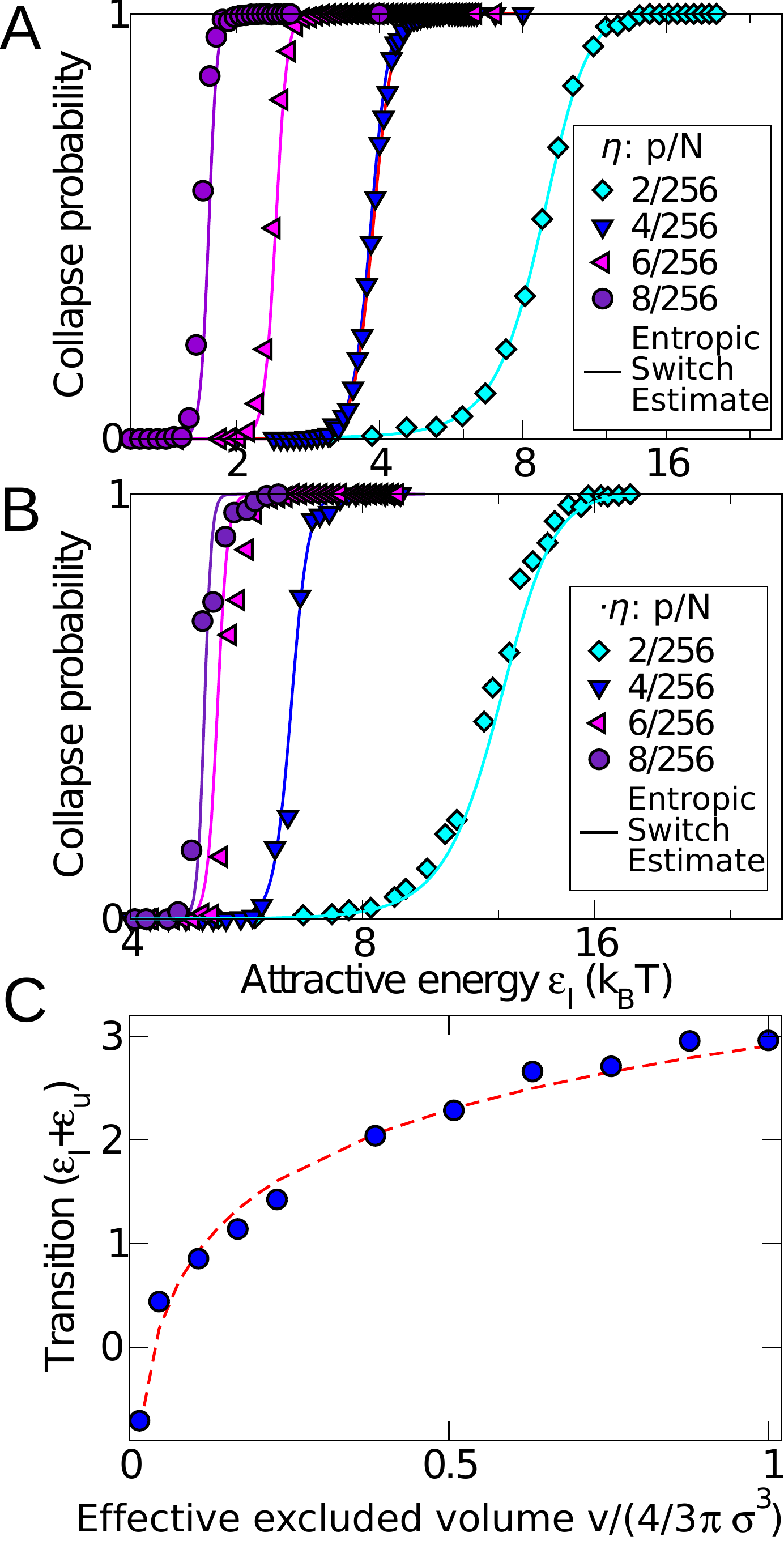}
  \caption{Role of entropy loss due to loop excluded volume
    interactions in the collapse due to sparse bridging. A: Collapse
    of a ghost polymer with bridging interactions. In this case, the
    inter-arms interactions are absent, and the plot shows good
    agreement of the predictions of Eqs.~\ref{eq:prob_collapseZ} and
    \ref{eq:Zloop_red} to simulation data. The formulas include a
    scaling of $P_\mathrm{loop} \sim g^{-3/2}$ measured from
    simulations. B: improved theoretical predictions by a shift in the
    transition energy of value $\Delta S_{star} = a_0 (2p -
    3/2)^{1.68} \log[256/p]$ (see Eq.~\ref{eq:grass}) where $ a_0 =
    0.00258$ is a fitted parameter. C: Agreement of the scaling of the
    transition point between the simulations for the polymer with
    bridging interactions ($p/N = 8/256$) in the swollen arm regime
    and the predictions of Eq.~\ref{eq:corona}.}
  \label{fig:6}
\end{figure}

This Appendix briefly addresses how accounting for loop interactions
can improve the estimates for the switch-like collapse presented in
Fig.~\ref{fig:3a}A.  The scaling of the probability of collapse for
swollen arms depends on the excluded volume interaction parameter $v$
through two effects.  Firstly, as already mentioned, and accounted in
the estimates, the swelling of the arms due to the self-interactions
changes the scaling for the contact probability from $P_\mathrm{loop}
\sim g^{-3/2}$ to $P_\mathrm{loop} \sim
g^{-2.27}$~\cite{Marenduzzo2006c}.  Secondly, and more importantly,
there is an interaction, so far neglected between the arms due to the
peculiar star-like configuration of the collapsed polymer.
The estimates of the switch-like transition from
Eqs.~\ref{eq:prob_collapseZ} and \ref{eq:Zloop_red} do not show
perfect agreement with simulations (Fig.~\ref{fig:3a}A) with increasing
$p$ because this entropy reduction due to the excluded volume
interactions between loops is neglected. This is confirmed by the fact
that the approximation of Eq.~\ref{eq:Zloop_red} (valid for
sufficiently small $p$) is very accurate for ghost chains
(Fig.~\ref{fig:6}A).

A loop interaction term must depends on the arm length $g$ as well as
from the interaction parameter $v$ and the number of arms $f$.  The
entropy reduction for a star polymer with $f$ arms of length $N/f$ has
been studied in the thermodynamic limit ($N \to
\infty$)~\cite{Schafer1992}. One can define a set of scaling exponents
$\gamma_f$ which define the scaling between the number of
configurations of a star polymer made of $f$ arms each of length $N/f$
as
\begin{equation}
  Z = \mu^{-N} \left(N/f\right)^{\gamma_f - 1} \ .
\end{equation}
Due to the effect of this scaling, the collapse energy of the polymer
with bridging interactions is shifted to higher values for increasing
$p$ by an entropic contribution as~\cite{Hanke2003}
\begin{equation}
\Delta S_{star} = \sigma_{2p} \log \frac{N}{p-1},\ 
\mathrm{with}\ \sigma_{2p} = \gamma_p - \frac{\gamma_1 + 1}{2} \ .
\label{eq:sstar}
\end{equation}
Numerical calculations of $\gamma_f$ for selected values of $f$ have
been carried out using field-theoretical methods~\cite{Schafer1992}
and lattice polymer
simulations~\cite{Batoulis1989,Grest1994,Caracciolo1998,Zifferer1999,Shida2000,
  Hsu2004}.

In a complementary way, the problem of the star polymer has been
approached by scaling arguments based on polymer
blobs~\cite{Daoud1982}. This approach shows that the scaling of the
free energy of the star polymer is proportional to $\sim
f^{3/2}$~\cite{Witten1986}, setting a scaling for $\gamma_f$. The
simulations used to compute $\gamma_f$ numerically gave~\cite{Hsu2004}
the relation
\begin{equation}
\gamma_f - 1 \simeq -(f-3/2)^{1.68}
\label{eq:grass}
\end{equation}
which is in agreement with the scaling estimate.

The simulations and scaling estimates discussed so far do not account
for the excluded volume strength. In order to include the role of
excluded volume, we propose a scaling for the entropic cost of the
star polymer term as in Eq.~\ref{eq:corona} (see
Appendix~B), which includes also the
logarithmic correction factor depending on the homogeneous interaction
term (and containing the parameter $v$ measuring interaction
strength).
Fig.~\ref{fig:6}B shows that adding such a term (scaling as
Eq.~\ref{eq:corona}) in Eq.~\ref{eq:Zloop_red} sensibly improves the
agreement of the theoretical estimate with the simulated switch-like
collapse.  Fig.~\ref{fig:6}C specifically tests the role of the
logarithmic dependency of the transition point from $v$, finding
satisfactory qualitative agreement.

\section*{Appendix B: Scaling argument for the entropy of a star polymer.}
\label{sec:scal-argum-entr}

These notes sketch the calculation of the number of blobs of a star
polymer with $f$ arms, each of length $N$, following Daoud and
Cotton~\cite{Daoud1982}. The number of blobs are then used as a proxy
for the entropy (blob ansatz).
The polymer is described as a series of concentric shells, each of
which by definition contains $f$ blobs (one for each arm). The size of
the blobs $\xi$ depends on the radial coordinate of the shell (called
$r$). Since the shell at $r$ has surface $\sim r^2$, the size of each
blob is $\xi^2 = r^2/f$, which means that $\xi \sim r / f^{1/2}$,
i.e. $\xi = a r$, with $a = C f^{-1/2} $.  $\xi$ is also the thickness
of the shell at coordinate $r$.

Starting form a core of size $r_0$ we now imagine to stack
(iteratively) discrete blobs of the proper size on each shell. Each
stacked shell will determine the coordinate of the following one, and
hence its blob size. 
\begin{displaymath}
  r_i = a \sum_{k=1}^{i-1} r_k + a \frac{r_i}{2} + a r_0 \ ,  
\end{displaymath}
with
\begin{displaymath}
  \xi_i = \xi(r_i) = a r_i \ ,
\end{displaymath}
%
%
leading to the expressions
\begin{displaymath}
  r_n \left(1 -\frac{a}{2} \right) = r_0 a (1+a)^{n-1} \ ,
\end{displaymath}
and
\begin{displaymath}
  \xi_n = r_0 a^2 (1+a)^{n-1} \ .
\end{displaymath}

We note now that the relationship $\xi_n \sim g_n^{3/5} v^{1/5} b$,
also has to hold (each blob is a swollen polymer of bond length $b$),
where $v=1/2 - \chi$, and $g$ is the number of monomers in a
blob. Hence,
\begin{displaymath}
  g_n = A e^{\frac{5}{3}\left(n-\frac{1}{2}\right)a} \ ,  
\end{displaymath}
where
\begin{displaymath}
 A= \frac{(a^2 r_0)^{5/3}}{v^{1/3} b^{5/3}} . 
\end{displaymath}

We now follow (on one branch of the star) all the monomers in all the
blobs, and impose that their total has to be $N$. Inverting this
relationship gives an estimate for $ \beta F \simeq f
N_{\mathbf{blob}}$ (since the argument involves the blobs of one
arm). One has
\begin{displaymath}
  \sum_{n=1}^{N_{\mathbf{blob}}} g_n = N \ ,
\end{displaymath}
hence
\begin{displaymath}
  N \simeq \int_{1}^{N_{\mathbf{blob}}} \mathrm{d}n  A e^{\frac{5}{3}n
    a} \ , 
\end{displaymath}
from which
\begin{displaymath}
  N_{\mathbf{blob}} \sim \frac{1}{a} \log N a / A
\end{displaymath}
where we neglected additive and multiplicative numerical constants,
leading to the expression
\begin{displaymath}
  \beta F \sim f^{3/2} \log 
  \left(  N f^{7/6} v^{1/3}  \left(\frac{b}{r_0} \right)^{5/3} \right) \ ,
\end{displaymath}
which is Eq.~\eqref{eq:corona}.




\end{document}